\documentstyle[epsfig,12pt]{article}

\newcommand\be{\begin{equation}}
\newcommand\ee{\end{equation}}
\newcommand\bea{\begin{eqnarray}}
\newcommand\eea{\end{eqnarray}}
\newcommand\bfi{\begin{figure}}
\newcommand\efi{\end{figure}}

\title{\begin{flushright}
{\normalsize NUC-MINN-00/05-T\\
February 2000 \\}
\end{flushright}
\vspace*{0.3in}
{\bf Modification of Z Boson Properties in Quark-Gluon Plasma}}
\author{{\bf J. I. Kapusta}$^{\dag}$ and
{\bf S. M. H. Wong}$^{\ddag}$\vspace*{0.1in}\\
 {\it School of Physics and Astronomy, University of Minnesota}\\
 {\it Minneapolis, MN 55455}}

\date{}

\begin{document}

\maketitle

\begin{center}
Abstract
\end{center}

We calculate the change in the effective mass and width of a Z boson in the 
environment of a quark-gluon plasma under the conditions expected in Pb-Pb 
collisions at the LHC.  The change in width is predicted to be only about 1 MeV 
at a temperature of 1 GeV, compared to the natural width of 2490$\pm$7 MeV.  The 
mass shift is even smaller.  Hence no observable effects are to be expected. 

\vspace*{1.in}

\vspace*{1.in} \noindent
$^{\dag}$kapusta@physics.spa.umn.edu\\
$^{\ddag}$swong@nucth1.hep.umn.edu
\newpage

The Large Hadron Collider (LHC) at CERN will collide lead nuclei at a
center-of-mass energy of 2.75 TeV per nucleon, more than an order of magnitude 
greater than the Relativistic Heavy Ion Collider (RHIC) at BNL.  At this energy 
Z bosons will be produced in measurable numbers. The mass and total width of the 
Z boson are $m_Z = 91.153 \pm 0.007$ and $\Gamma_{\rm vac} = 2.490 \pm 0.007$ 
GeV, respectively \cite{pdg}.  Kinematics and phase space then dictate that they 
will normally be produced with small transverse velocities and, as we shall see, 
most will decay in a dense environment of quarks and gluons.  How much will 
their mass and width be modified due to immersion in this quark-gluon plasma, 
and how can these effects be measured?

The standard picture of a central collision between high energy heavy nuclei is 
the boost invariant hydrodynamic model of Bjorken \cite{bj}.  Following an 
initial pre-equilibrium phase, quarks and gluons are approximately 
thermalized in the local rest frame at proper time $t_0$ after the highly 
Lorentz contracted nuclei overlap with an initial temperature $T_0$.  Based on 
very general considerations these numbers were estimated to be 0.07 fm/c and 1 
GeV, respectively, for central Pb-Pb collisions at the LHC \cite{big}.  
Thereafter the local temperature falls with proper time as
\be
T(t) = \left(\frac{t_0}{t}\right)^{1/3} T_0 \, .
\ee
The quark-gluon plasma expands longitudinally until it hadronizes when the local 
temperature reaches $T_c \approx 160$ MeV.  After that, transverse expansion 
sets in.  Eventually the hadrons lose thermal contact and begin a free-streaming 
phase.  By the uncertainty principle a Z boson is created approximately $1/m_Z = 
0.002$ fm/c after nuclear contact, and it decays with a (vacuum) lifetime of 
0.08 fm/c.  Therefore, Z bosons will decay in an environment of quark-gluon 
plasma with a temperature of order 1 GeV and so have the potential to be an 
excellent probe of the highest temperatures attained in these collisions. 

The spectral density of the Z boson evaluated at small velocity $v \ll c$ in the 
rest frame of the plasma is
\be
\rho(s,T) = \frac{1}{\pi} \frac{m_Z \Gamma}{(s-m_Z^2)^2 + m_Z^2 \Gamma^2}
\ee
where $\Gamma = \Gamma_{\rm vac} + \Gamma_{\rm mat}(T)$, the latter contribution 
denoting the effect due to matter, namely the quark-gluon plasma.  The invariant 
mass distribution of dimuons coming from Z decay is directly proportional to the 
spectral density.   Given that the Z bosons are invariably created at time 
$1/m_Z \equiv t_Z$ after nuclear overlap, the number of them remaining at time 
$t > t_Z$ is
\be
N_Z(t) = N_Z(t_Z) \exp\left[ - \int_{t_Z}^t \Gamma(t') dt' \right] \, .
\ee
Here $\Gamma$ will depend on the time because it depends on temperature and that 
in turn depends on time.  The final observed distribution of dimuons coming from 
Z decay will be
\be
\frac{dN_{\mu^+ \mu^-}}{ds} = \Gamma_{\mu^+ \mu^-} \int_{t_Z}^{\infty} dt N_Z(t) 
\rho(s,T(t)) \, .
\ee
At the LHC the lifetime of the Z closely matches the thermalization time $t_0$.  
Therefore it is a good approximation to evaluate the spectral density at this 
time, resulting in the distribution
\be
\frac{dN_{\mu^+ \mu^-}}{ds} = \frac{\Gamma_{\mu^+ \mu^-}}{\Gamma(T_0)} N_Z(t_Z) 
\rho(s,T_0) \, .
\ee
The total number of Z bosons $N_Z(t_Z)$ ought to scale as $A^2$ for impact 
parameter averaged collisions ($A$ is the atomic number) because production of a 
Z is such a hard process.  A measurement of the dimuon mass spectrum is being 
planned by the CMS \cite{cms} collaboration at the LHC.  How large might one 
expect the many-body effects to be?

\bfi
\centerline{\epsfig{figure=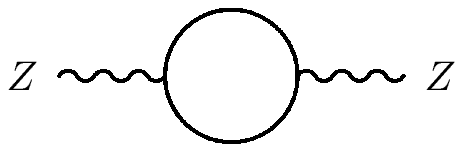,width=2.0in}}
\caption{One-loop contribution to the Z self-energy at finite 
temperature. The solid line represents a light quark (u,d,s).}
\label{fig1}
\efi

The position of the pole of the Z boson propagator at finite temperature may be 
computed with standard techniques. The lowest order correction to the real part 
of the Z self-energy comes from the single quark loop diagram shown in Fig. 
\ref{fig1}. In the limit that the Z moves non-relativistically in the plasma the  
longitudinal and transverse self-energies are degenerate.  The real part is (we 
use the conventions of Cheng and Li \cite{ChengLi})
\bea
{\rm Re} \, \Pi_{\rm 1-loop} &=& -\frac{14 \pi^2}{15} \frac{T^4}{m_Z^2}
\sum_f \left[g_A^2(f) + g_V^2(f)\right] \nonumber \\
&=& -\frac{7\sqrt{2}}{5} \pi^2 \left[ 1-\frac{4}{9} \sin^2(2\theta_W) \right]
G_F T^2 \, .
\eea
Here it is assumed that $T \ll m_Z$.  The summation is over the relevant flavors 
of quarks.  The second line includes only u, d and s quarks.  Using the latest 
values \cite{pdg} for the Fermi constant, $G_F = 1.166 \times 10^{-5}$ GeV$^{-
2}$, and weak angle, $\sin^2\theta_W = 0.2312$, the mass shift is
\be
{\rm Re} \, \Pi_{\rm 1-loop} = \Delta m_Z^2 = -1.56\times10^{-4} \, 
T^4_{\rm GeV}
\ee
with temperature measured in GeV.  The mass shift is negative but 
uninterestingly small.  The finite temperature imaginary part of the one-loop 
diagram simply amounts to a Pauli blocking factor for the quarks in the final 
state.  This is $2 \exp(-m_Z/2T) \approx 3\times 10^{-20}$ for $T = 1$ GeV which 
is totally irrelevant.

\bfi
\centerline{\epsfig{figure=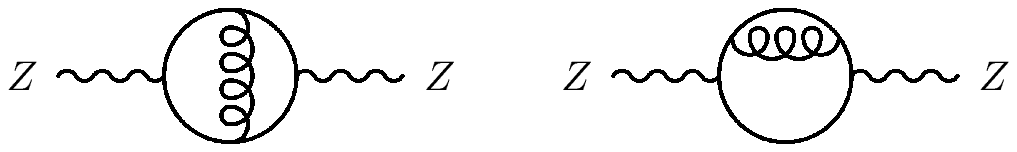,width=4.20in}}
\caption{Two-loop contributions to the Z self-energy at finite temperature.  
The solid line represents a light quark (u,d,s) and the curly line represents 
a gluon.}
\label{fig2}
\efi

%\vspace{-2.0cm}

\bfi
\centerline{\epsfig{figure=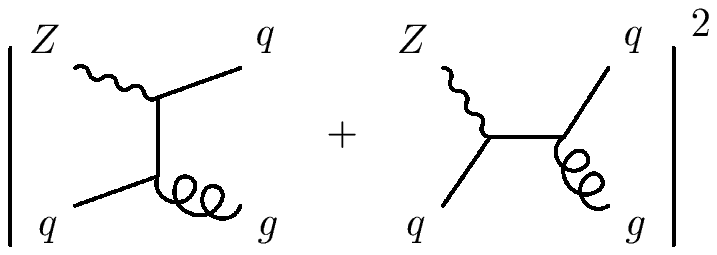,width=2.50in}}
\vspace{0.3cm}
\centerline{\epsfig{figure=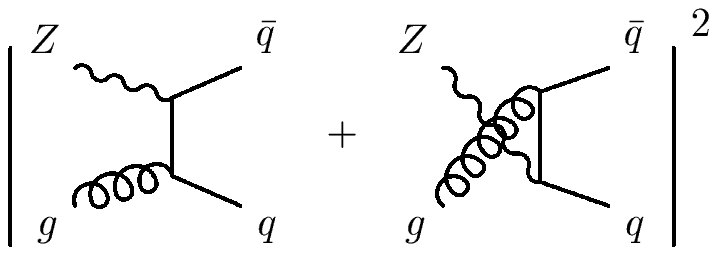, width=2.50in}}
\vspace{0.3cm}
\centerline{\epsfig{figure=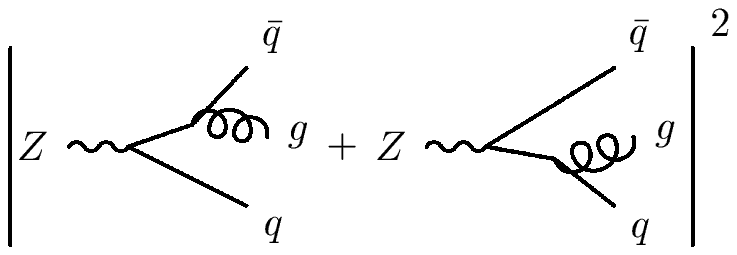,  width=2.50in}}
\vspace{0.3cm}
\centerline{\epsfig{figure=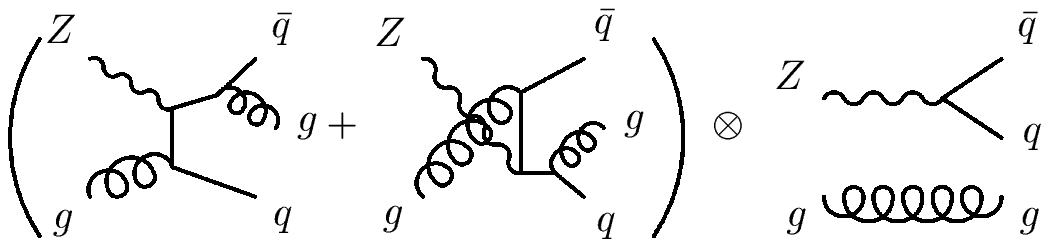,   width=3.70in}}
\caption{Cutting the two-loop self-energy in Fig. 2 results in 
these contributions to the imaginary part: $q+Z \rightarrow q+g$ (Compton), 
$g+Z \rightarrow q+\bar{q}$ (fusion), $Z \rightarrow q+\bar{q}+g$ (3-body 
decay). There are also interferences between the amplitudes for $g+Z \rightarrow 
q+\bar{q}+g$ and $Z \rightarrow q+\bar{q}$ together with a spectator gluon, an 
example of which is shown here (the incoming and outgoing gluons have the same 
energy and momentum).  There are similar interference terms involving a 
spectator quark or antiquark.} 
\label{fig3}
\efi

The dominant finite temperature effect on the imaginary part of the 
self-energy comes from the two-loop diagram shown in Fig. \ref{fig2}. 
Cutting that diagram in all possible ways corresponds to scattering 
processes some of which are shown in Fig. \ref{fig3}. They represent 
the reactions $q + Z \rightarrow q + g$ (Compton), 
$g + Z \rightarrow q + \bar{q}$ (Fusion), $Z \rightarrow g + q + \bar{q}$ 
(3-body Decay), an interference between the amplitudes for 
$g + Z \rightarrow g + q + \bar{q}$ and $Z \rightarrow q + \bar{q}$ with 
a spectator gluon (Interference/Gluon), and an interference between the 
amplitudes for $q(\bar{q}) + Z \rightarrow q({\bar q}) 
+ q + \bar{q}$ and $Z \rightarrow q + \bar{q}$ with a spectator quark or 
antiquark (Interference/Quark). Averaged over initial spins and 
summed over final spins and color the invariant amplitudes are:
\be
|{\cal M}_C|^2 \propto - \frac{s}{u} - \frac{u}{s}
-2t \left(\frac{1}{s} + \frac{1}{u} + \frac{t}{su} \right)
\ee
\be
|{\cal M}_F|^2 \propto  \frac{t}{u} + \frac{u}{t}
+2s \left(\frac{1}{t} + \frac{1}{u} + \frac{s}{tu} \right)
\ee
\be
|{\cal M}_D|^2 \propto 2\left( \frac{p\cdot p'+k\cdot p'}
{p\cdot q} + \frac{p\cdot p'+k\cdot p}{ p'\cdot q}
+ \frac{2(p\cdot p')^2}{(p\cdot q)( p'\cdot q)} \right) \, .
\ee
In all cases the constant of proportionality is $16 g_s^2 (g_V^2+g_A^2)/3$.  The 
four-momenta are $p$ for quark, $p'$ for antiquark, $q$ for gluon and $k$ for 
the Z boson.  The contribution to the imaginary part of $\Pi$ involves 
integrating these amplitudes over phase space in the usual way, including 
thermal distributions of quarks and gluons in the initial state and Pauli 
blocking or Bose enhancement factors in the final state.  The interference
terms are more complicated and are not displayed here.  A detailed exposition
of how they are derived and evaluated in this and other theories is deferred
to a separate paper \cite{comment}.

The contribution to the imaginary part of $\Pi$ involving one gluon thermal 
distribution, arising from the Fusion, Decay and Interference processes may be 
written as (limits $m_f \ll T \ll m_Z$ assumed throughout)
\be
{\rm Im} \, \Pi_{F+D+I} = -\frac{2}{3} \frac{\alpha_s}{\pi^2}    
\sum_f \left[g_A^2(f) + g_V^2(f)\right] \int_0^{\infty} d\omega
\frac{1}{\exp(\omega/T)-1} \left( F+D+I \right)
\ee
where
\bea
F &=& -2\omega +\left[ 2\omega +2m_Z +\frac{m_Z^2}{\omega} \right]
\ln\left( \frac{2m_Z \omega}{k_c^2} \right) \, , \\
D &=& -2\omega +\left[ 2\omega - 2m_Z +\frac{m_Z^2}{\omega} \right]
\ln\left( \frac{2m_Z \omega}{k_c^2} \right) \, , \\
I &=& 8\omega +\left[ 4\omega -2\frac{m_Z^2}{\omega} \right]
\ln\left( \frac{2m_Z \omega}{k_c^2} \right) \, .
\eea
Here $\omega$ is the energy of the gluon and $k_c^2$ is a cutoff placed on the 
four momentum transfer $t$ and/or $u$ (or virtuality)
carried by the exchanged quarks in the plasma \cite{photon}.  It arises 
naturally in the infrared resummation scheme of Braaten and Pisarski 
\cite{resum}.  The 
individual contributions are quadratically divergent in the infrared.  The sum 
of the three contributions is, however, finite as expected by rather general 
arguments \cite{art}.
\be
{\rm Im} \, \Pi_{F+D+I} = -\frac{1}{9} \alpha_s 
\sum_f \left[g_A^2(f) + g_V^2(f)\right] 
T^2 \left[ 8\ln\left( \frac{2 m_Z T}{k_c^2} \right) + 2.8224 \right]
\ee

The contribution from processes involving one quark or antiquark thermal 
distribution may similarly be computed to be
\be
{\rm Im} \, \Pi_{C+D+I} = -\frac{2}{3} \frac{\alpha_s}{\pi^2}    
\sum_f \left[g_A^2(f) + g_V^2(f)\right] \int_0^{\infty} d\omega
\frac{1}{\exp(\omega/T)+1} \left( C+D+I \right)
\ee
where
\bea
C &=& 4\omega + \left[ -2\omega + m_Z \right] 
\ln\left( \frac{2m_Z \omega}{k_c^2} \right) \, ,  \\
D &=& 2\omega +\left[ - 2\omega - m_Z \right]
\ln\left( \frac{2m_Z \omega}{k_c^2} \right) \, ,  \\
I &=&  8\omega \ln\left( \frac{2m_Z \omega}{k_c^2} \right) \, .
\eea
After integration, the sum of these contributions is
\be
{\rm Im} \, \Pi_{C+D+I} = - \frac{1}{9} \alpha_s 
\sum_f \left[g_A^2(f) + g_V^2(f)\right] 
T^2 \left[ 2 \ln\left( \frac{2 m_Z T}{k_c^2} \right) + 4.0920 \right] \, .
\ee 
The sum of all processes, F+C+D+I, is
\be
{\rm Im} \, \Pi_{F+C+D+I} = - \frac{10}{9} \alpha_s 
\sum_f \left[g_A^2(f) + g_V^2(f)\right] 
T^2 \left[ \ln\left( \frac{2 m_Z T}{k_c^2} \right) + 0.6914 \right] \, .
\ee

One can view the cutoff on the quark virtuality as the effective mass of a quark 
propagating through the plasma with a typical thermal momentum.  This effective 
mass is $\sqrt{2}m_q$ where $m_q^2 = g_s^2 T^2 /6$ \cite{photon}.  Summing over 
the three flavors of lightest quarks,
\be
{\rm Im} \, \Pi = - \sqrt{2} \left[ 1-\frac{4}{9} \sin^2(2\theta_W) \right] 
\alpha_s 
m_Z^2 G_F T^2 \ln\left( \frac{m_Z}{1.049 \alpha_s T} \right) \, ,
\ee       

The rate $\Gamma$ is determined from the self-energy via $\Gamma = -$Im 
$\Pi/m_Z$.  Numerically the thermal contribution to the rate is
\be
\Gamma_T = 1.03\times 10^{-3} \alpha_s T^2_{\rm GeV}
\ln\left( \frac{86.9}{\alpha_s T_{\rm GeV}} \right) \, .
\ee
For the strong coupling we use the one loop expression 
\be
\alpha_s(T) = \frac{6\pi}{27 \ln(T/50 \, {\rm MeV})} \, ,
\ee
once again assuming 3 flavors of massless quarks.  The argument of the logarithm 
is adjusted to fit lattice results \cite{coupling}.  Numerically
$\alpha_s(T = 200$ MeV) $\approx$ 1/2 and $\alpha_s(T = 1$ GeV) $\approx$ 1/4.  
At $T = 1$ GeV this gives a finite temperature contribution to the width of only 
1.5 MeV and totally ignorable compared to the vacuum part.

In conclusion, we have computed the finite temperature contribution to the width 
of the Z boson in a quark-gluon plasma achievable in Pb-Pb collisions at the 
LHC.  This effect will not be observable in the dimuon invariant mass spectra. 
This is very unfortunate because the approximate equality between the Z lifetime 
and the formation/thermalization time of the quark-gluon plasma could have made 
it a fine probe of the initial state of the plasma.

\section*{Acknowledgments}

We thank P. Stankus and R. Rusack for prodding us to investigate this problem.  
This work was supported by the US Department of Energy under grant
DE-FG02-87ER40328.

%\newpage

%\section*{Figure Captions}

%\noindent{\bf Figure \ref{fig1}:} One-loop contribution to the Z self-energy 
%at finite temperature.  The solid line represents a light quark (u,d,s).\\

%\noindent{\bf Figure \ref{fig2}:} Two-loop contributions to the Z self-energy at 
%finite temperature.  The solid line represents a light quark (u,d,s) and the 
%curly line represents a gluon.\\

%\noindent
%{\bf Figure \ref{fig3}:} Cutting the two-loop self-energy in Fig. 2 results in 
%these contributions to the imaginary part: $q+Z \rightarrow q+g$ (Compton), $g+Z 
%\rightarrow q+\bar{q}$ (fusion), $Z \rightarrow q+\bar{q}+g$ (3-body decay).
%There are also interferences between the amplitudes for $g+Z \rightarrow 
%q+\bar{q}+g$ and $Z \rightarrow q+\bar{q}$ together with a spectator gluon, an 
%example of which is shown here (the incoming and outgoing gluons have the same 
%energy and momentum).  There are similar interference terms involving a 
%spectator quark or antiquark. 

\end{document}